\documentclass[prb,twocolumn,showpacs]{revtex4}
\usepackage{graphicx}
\usepackage{amsmath}
\usepackage{color}

\begin{document}

\title{Modeling of microwave-assisted switching in micron-sized magnetic ellipsoids}

\author{R. Yanes}
\author{R. Rozada}
\author{F. Garc\'ia-S\'anchez}
\author{O. Chubykalo-Fesenko}

\affiliation{ Instituto de Ciencia de Materiales de Madrid, CSIC,
Cantoblanco, 28049 Madrid, Spain}

\author{P. Martin Pimentel}
\author{B. Leven}
\author{B. Hillebrands}

\affiliation{ Fachbereich Physik and Research Center OPTIMAS, Technische Universit\"{a}t Kaiserslautern, Erwin-Schr\"{o}dinger-Str. 56, 67663 Kaiserslautern, Germany}

\begin{abstract}
The microwave assisted magnetisation reversal is modelled in a permalloy micron-sized magnetic ellipsoid. Our simulations confirm that this process requires less field than magnetisation reversal under a static field. This is due to a different reversal mode which in case of the microwave-assisted process is always a ripple structure. During the magnetisation reversal two stages: nucleation and relaxation are distinguished. The nucleation process is governed by spinwave instabilities. The relaxation process is related to the domain expansion through domain wall propagation determined by the precessional motion of magnetic moments in the center of the domain walls. As a consequence, the switching time is a complex oscillating function of the microwave frequency.

\end{abstract}


\maketitle

\section{Introduction} 
The fundamentals of fast magnetisation switching constitute one of the important problems in magnetism, related to technological applications \cite{Hillebrands}. Experimentally  microwave-assisted switching has been reported as requiring a smaller applied field thus opening new possibilities for applications in magnetic recording and communication devices \cite{Thirion, Nembach, Woltersdorf,Moriyama, Pimentel}. The microwave-assisted (mw) switching process with linearly-polarized excitation can be realized based, for example, on a fast magneto-optic Kerr effect set-up, applying simultaneously static and microwave fields \cite{Nembach,Woltersdorf, Pimentel, Adam}. Another possibility to observe  mw-assisted switching is given by measuring the magnetoresistance as has been done, for example, in Co stripes \cite{Grollier} or in NiFe magnetic tunnel junctions \cite{Moriyama}. These experiments use relatively large (up to tens of microns) lithographically prepared magnetic elements whose dimensions do not allow the occurrence of homogeneous magnetisation processes.

 Differently from this, the micro-SQUID experiment \cite{Thirion}, performed on Co nanoparticles with 20 nm diameter gives an example of a macrospin magnetisation behavior. In this system nonlinear effects are clearly observed experimentally in agreement with a  simple macrospin dynamical model.
  The behavior found in the macrospin magnetisation dynamics has stimulated the technologically important proposal of micro-wave assisted (or ferromagnetic resonance (FMR) assisted) magnetic recording \cite{Zhu, Rivkin, Scholtz}.

 The process of understanding the underlying physics is still going on and several ideas are found in the literature. From the theoretical point of view, probably the most comprehensive understanding has been achieved in the case of one single magnetic moment \cite{Thirion, Rivkin, Scholtz, Sun, Zhang, Lee, Horley}.
 In what follows we would like to mention some of the proposed mechanisms, since most of them can also be partially encountered in the complex situation of non-homogeneous magnetisation reversal processes discussed in the present article. The most frequently used one is based on a simple idea of the efficient energy transfer during excitation with frequencies close to the FMR one \cite{Woltersdorf, Sun}. It has been also argued that excitation with a circular polarised field would be much more efficient due to the possibility to synchronize with the rotation sense of the magnetic moment \cite{Sun}. Although it seems to be less efficient,  synchronization is also possible with a linearly polarised field \cite{Horley}. This idea remains certainly valid in a general case, however it should be noted that the notion of the FMR frequency, as corresponding to excitation of mode with zero wave vector, is not exact for micron-sized magnetic elements. Due to the minimization of the magnetostatic charges at the surfaces, even small angle magnetisation vibrations are not homogeneous \cite{Adam} and the frequency spectrum depends strongly on the size and shape of the magnetic nanoelement \cite{Kruglyak1, Kruglyak}.
 Moreover, in magnetic nanoelements the frequency spectrum becomes quantized \cite{Bayer,Guslienko}. The second idea found in the literature is also simple:
 The microwave field provides a constant energy input which rises the magnetic energy of the particle up to a level higher than the saddle point of the magnetic energy landscape \cite{Thirion, Sun, Moriyama, Horley}. Note that here also the thermal-activated process becomes possible \cite{Lee,Nozaki}. In this sense, it can be said that the microwave field can act as an energy source which effectively decreases the energy barrier. It should be noticed here that this idea is relied on a simple picture of a two-level system which is generally not valid for a non-homogeneous magnetisation reversal scenario.

The fast mw-assisted magnetisation switching process in magnetic elements is closely related to the phenomenon of precessional switching. Indeed, a small perpendicular field helps the switching to take place via the excitation of the precessional motion \cite{Back, BackScience, Hiebert} due to the torque acting on the magnetisation. The main problem is that with, for example, a perpendicularly applied field the magnetisation continues to precess ("ringing" phenomenon) and could switch back \cite{Schumacker}. Furthermore, it has been suggested to use a pulsed field and to tune the pulse duration to avoid multiple switching. Even the pulse shape could be optimized \cite{Bauer}, together with the use of two pulses with an optimised delay time \cite{Gerrits, Schumacker}. This case again is mostly studied in the case of one magnetic moment, although micromagnetic simulations have been also widely performed in nanoelements with different shapes \cite{Viena}.

The final idea which we would like to mention is that one magnetic moment under a microwave field represents a classical example of a parametrically excited nonlinear oscillator. Thus the problem is similar to the classical parametrically excited pendulum, where  parametric instabilities occur varying the strength and the frequency of external oscillation. Such instabilities may be considered as precursors of the switching process. Next, in these systems nonlinear phenomena such as bifurcations and chaos \cite{Pla, Serpico, Horley} occur. A classical example of this is the appearance of an additional large-amplitude stationary orbit via the folding  bifurcation, again studied in the case of one macrospin only \cite{Pla, Serpico}.  The nonlinear phenomena lead to a complex structure of the trajectories  near the separatrix, a  special trajectory separating the basis of attractions of the two minima. Under the influence of external periodic force, the separatrix may become fractal so that closely situated initial conditions may lead either to switching or not \cite{Horley}. This produced, for example, fractal regions of switching beyond the Stoner-Wolfarth astroid observed experimentally in micro-SQUID experiments \cite{Thirion} and numerically in Refs.\onlinecite{Thirion, Zhang, Horley, Scholtz}.  However, since these phenomena are complex even in the case of one magnetic moment, in a large system  nonlinear phenomena are more difficult to interpret and may not lead to a clear pattern.

The reversal under oscillating fields of large-sized magnetic nanoelements was mostly studied in  cases with simple magnetisation reversal modes, such as domain wall\cite{Grollier,Martinez}, $C$ or $S$ magnetisation states \cite{Fidler}, the vortex to onion state in magnetic rings \cite{Podbielski} or vortex core reversal \cite{Wayenberge, LeeK}. The enhanced mobility of domain walls with some frequencies was numerically found \cite{Martinez}. Experimentally, using the magnetic circular dichroism technique in a magnetic element having a Landau pattern, the asymmetry of the domain wall motion under a linearly-polarised mw field with the drift in one direction was reported \cite{Krasyuk}. This has been explained by the entropy maximization involving the precessional excitation (the excitation of a so-called "self-trapping spin-wave mode").

As we mentioned above, most of the previous theoretical studies involve either fast switching in the macrospin case (or effectively conditions of coherent rotation) or the switching of individual objects as vortices or domain wall structures. However, the spatial resolution of the fast Kerr experiments \cite{Nembach, Pimentel} requires quite large magnetic elements with not so simple and unique magnetisation reversal modes. Namely, the mw-assisted switching in large systems produces a sequence of nucleation - propagation - relaxation phenomena which is the subject of the present study. We show that the overall dynamics is a complex phenomenon where some of the above mentioned  processes could occur simultaneously and could play a role on different stages of the magnetisation reversal.

\section{Model}
\begin{figure}[h!]
\includegraphics[height=6cm]{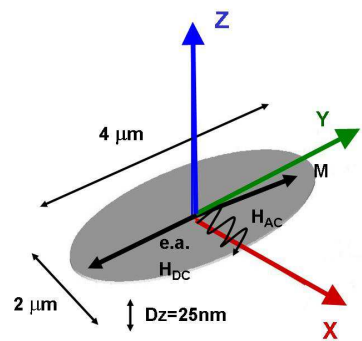}
\caption{Geometry of the simulated system.}
\label{geometry}
\end{figure}

In the present work and with the aim to understand the mw-assisted switching processes in large magnetic elements \cite{Nembach,Pimentel}, we use a micromagnetic model to simulate the magnetisation dynamics in a permalloy ellipsoid. Due to the limitation of the simulational size we used  reduced dimensions of the ellipsoid:  the total magnetic volume is $V= 4 \mu m$ x $2 \mu m$ by $25 nm$. It has the same aspect ratio as in the experiment. We used the publicly available code Magpar \cite{Magpar} with a finite element discretisation and in which we implemented the possibility to simultaneously apply dc and ac fields.
The permalloy easy axis was directed parallel to the large ellipsoid axis (y-axis, see Fig.~\ref{geometry}). Initially the ellipsoid was magnetised along this axis. The dc-applied field is placed parallel to it in the opposite direction. The linearly polarized ac field with large amplitude is directed perpendicular to it (x-axis). The following micromagnetic parameters were used in the simulations:  anisotropy value $K=3.31\times 10^3 erg/cm^3$, saturation magnetisation value $M_{s}=860 \;emu/cm^3$ (the anisotropy field $H_{K}=7.7 Oe$),  exchange parameter $A=1.05 \times 10^{-6} erg/cm$,  damping parameter $\alpha=0.012$.

\begin{figure}[h!]
\includegraphics[height=6cm]{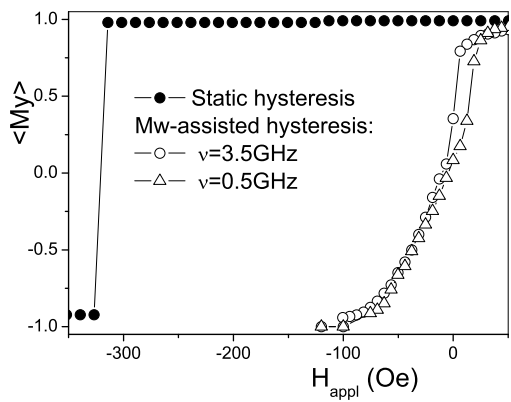}
\caption{Simulated descending branch of the hysteresis cycle for the permalloy ellipsoid with static applied field only (full circles) and with simultaneously applied static and mw field with $H_{ac}=25.1 Oe$, and two frequencies (open symbols). The dc-field is applied along the long ellipsoid axis $Y$. The average ellipsoid magnetisation $<M_y>$  is normalized to the total saturation value.}
\label{Hyst}
\end{figure}

In Fig. \ref{Hyst} we present a hysteresis cycle for the permalloy ellipsoidal element with the field applied parallel to the easy axis direction. The coercive field is much larger than the anisotropy field value due to the large magnetostatic shape anisotropy.

\section{Results}
First, we model the hysteresis cycle with simultaneously applied dc and ac fields (see Fig.~\ref{Hyst}) and note the reduction of coercivity  in the case of the mw-assisted switching process. The hysteresis cycle is square-like in the former case and rounded in the latter one. To understand the coercivity reduction we model the minimum strength of the applied field required to produce the magnetisation reversal in our system for different mw-field frequencies.

\begin{figure}[h!]
\includegraphics[height=6cm]{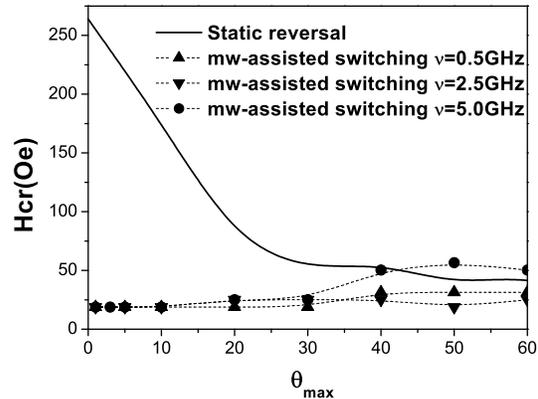}
\caption{Amplitude of the critical field necessary for magnetisation switching as a function of the maximum field angle with the anisotropy axis for static field and microwave-assisted switching. }
\label{DC_AC}
\end{figure}

We notice that one of the large contributions to the coercivity reduction comes from the deviation of the magnetisation direction from the y-axis, following the ac-field. Thus, the coercivity is reduced  simply due to the fact that most of time the resulting field is applied at some angle to the easy axis. Consequently, we compare the mw-assisted switching process with the situation when a static field is applied at an angle equal to that formed during the mw-assisted case with the maximum amplitude $H_{ac}$ of the ac field. The maximum amplitude of the critical field $H_{cr}=\sqrt{H_{ac}^2+H_{dc}^2}$, necessary to switch the magnetisation, is presented in Fig.~\ref{DC_AC} as a function of the maximum applied field angle. We use the following condition for the magnetisation switching $<M_y> < 0$, where $<M_y>$ is the average magnetisation along the $y$ direction. Our results show that even with this more fair comparison,  mw-assisted switching always requires less field, except for the case of very large deviation angles and some frequencies. Therefore, the first important contribution to decrease the coercivity during the mw-assisted switching process is the deviation of field from the easy axis which leads to precessional switching. Moreover, for small field angles the results are almost independent of the ac-field frequency which role is just to induce the precession. The results are different when strong deviations of the magnetisation occur which stresses the importance of the nonlinear effects.

\begin{figure}[h!]
\includegraphics[height=5cm]{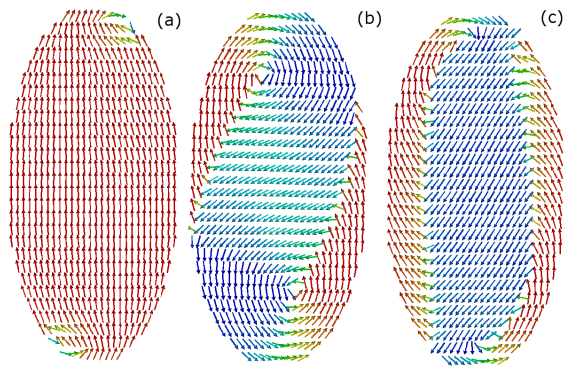}
\caption{Magnetisation configurations during the static field hysteresis process at different time moments  (a) initial state at the coercive field, (b) intermediate state, showing the vortices propagation and
(c)the final state at the coercive field. The field is applied at $45^o$ with the $Y$ axis.}

\label{Static}
\end{figure}

\begin{figure}[h!]
\includegraphics[height=6cm]{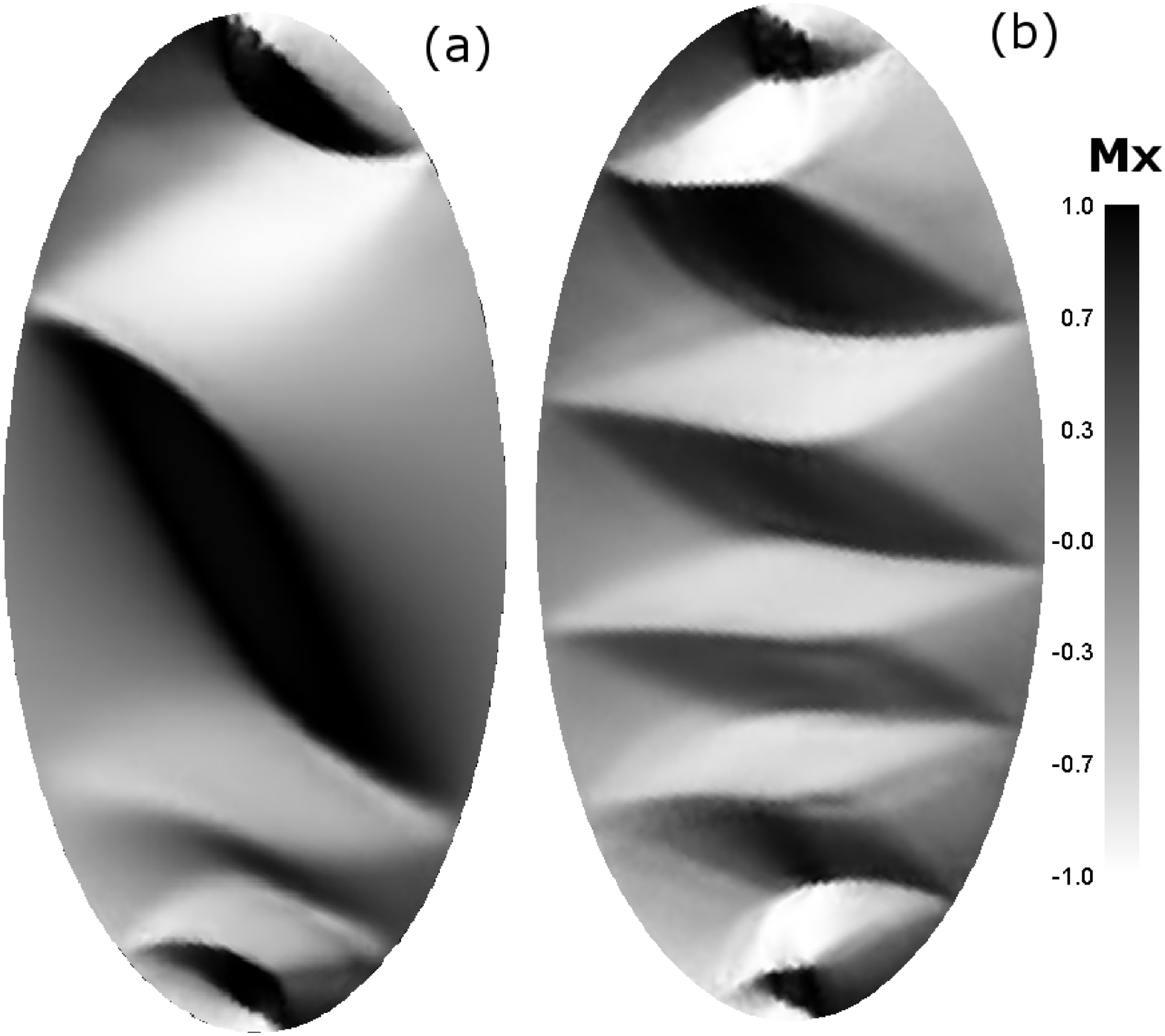}
\includegraphics[height=6cm]{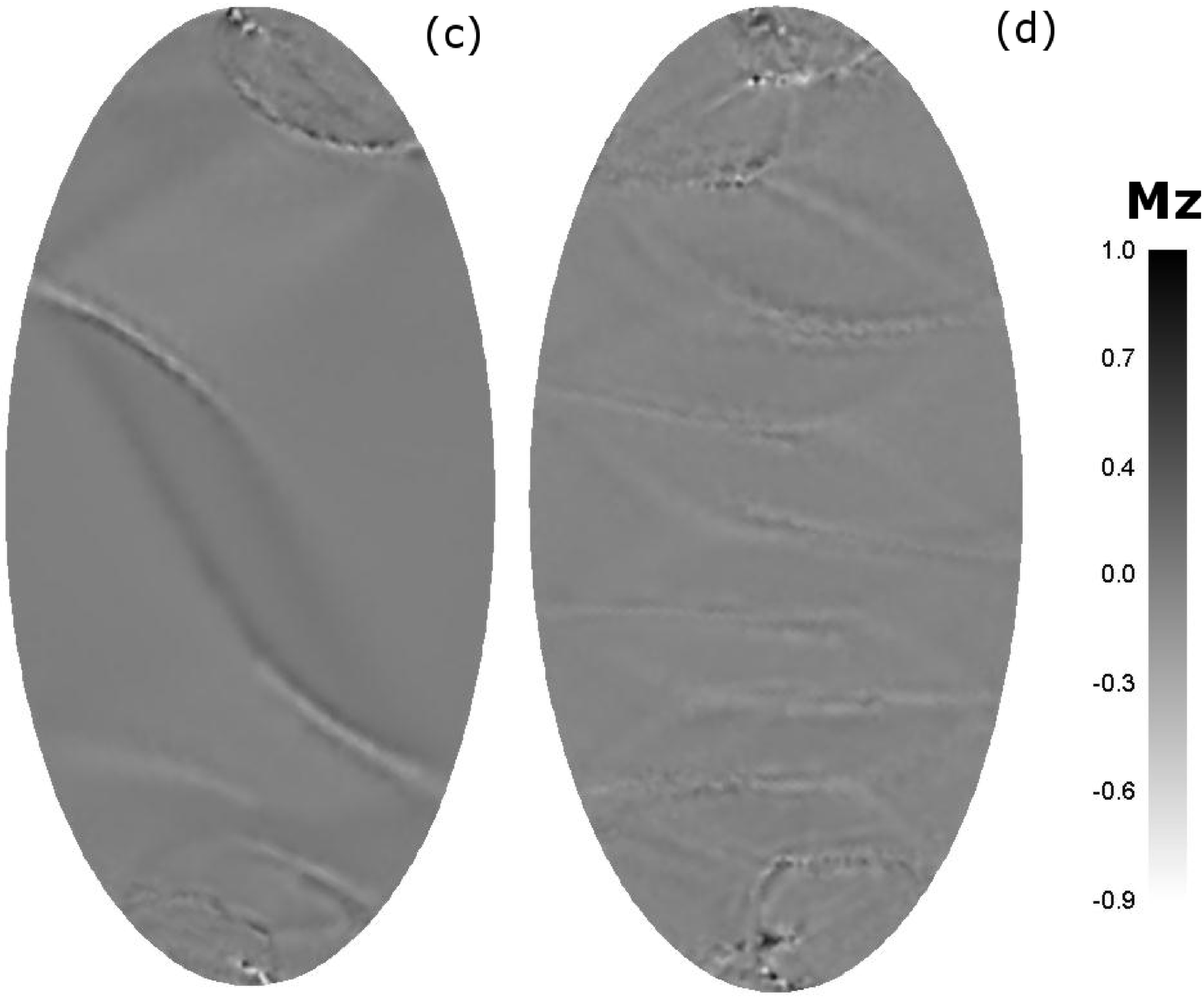}
\caption{Dynamical configurations: $M_x$ (upper) and $M_z$ (lower) components (grey scale) during the mw-assisted magnetisation reversal for applied fields $H_{dc}=-31.41 Oe$, $H_{ac}=18.84 Oe$ and for two different frequencies (left) $\nu=1 GHz$ and (right) $\nu=6 GHz$. For the dynamics of the reversal process, see the supplemental material \cite{Videos}.}
\label{MW}
\end{figure}

To illustrate the differences between the static and mw-assisted switching processes we present the dynamical configurations during the magnetisation reversal in Figs.~\ref{Static} and~\ref{MW}. For the static process (Fig.~\ref{Static}) the magnetisation nucleation starts with two vortices in the opposite upper and lower sides of the ellipsoid. During the irreversible jump these vortices propagate in  opposite directions creating two domain walls. The propagation of these domain walls toward the left and the right sides of the ellipsoid completes the magnetisation reversal.

\begin{figure}[h!]
\includegraphics[height=6cm]{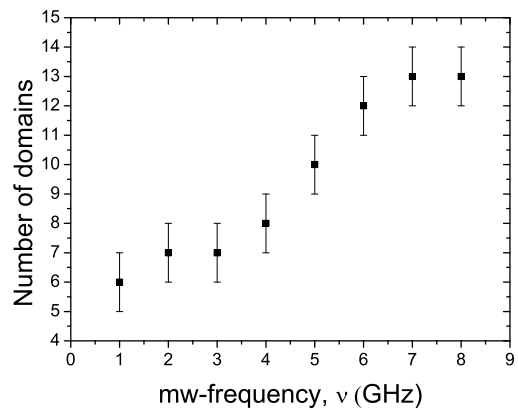}
\caption{ Number of domains in the ellipsoid nucleated during the first two nanoseconds as a function of mw frequency for applied field $H_{dc}=-31.41 Oe$, $H_{ac}=25.13 Oe$.}
\label{ripple}
\end{figure}

\begin{figure}[h!]
\includegraphics[height=6cm]{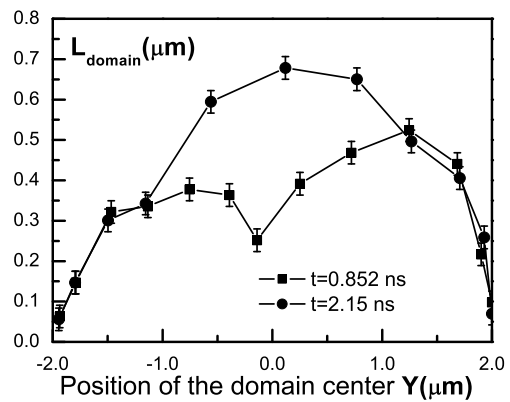}
\caption{Domain size in the ellipsoid along the $y$ coordinate as a function of the position of its center for two time snapshots, corresponding to the first nucleation-expansion stage of the mw-assisted reversal processes and for $H_{dc}=-31.41 Oe$, $H_{ac}=25.13Oe$. Note that after approximately $t> 2.5 ns$ the domain size expansion is suppressed.}
\label{size}
\end{figure}

Fig.~\ref{MW} shows the magnetisation configurations during the mw-assisted switching process. The complimentary video material (see Ref.~\onlinecite{Videos}) illustrates the temporary dynamical evolution of the demagnetisation reversal. We observe that the ellipsoid is dynamically divided into domain structures. The number of domains depends strongly on the mw frequency and may be related to the length of the spinwave mode, excited in the system. In Fig.~\ref{ripple} we represent an approximate number of magnetisation domains nucleated during the first two nanoseconds as a function of the ac-field frequency and in the magnetisation range $0.6 <\; <M_y> \; < 0.8$ (normalized to the total saturation magnetisation). Note that the domain size is larger in the central part of the ellipsoid than in the upper and lower sides, due to the ellipsoidal form, and initially grows with time (see Fig.~\ref{size}).
The division of the ellipsoid into domains during mw-assisted switching is experimentally confirmed by Kerr images in Ref.~\onlinecite{Pimentel}. Therefore, the mw-assisted case is characterized by a reversal mode completely different from the one appearing during the static hysteresis.

\begin{figure}[h!]
\includegraphics[height=5cm]{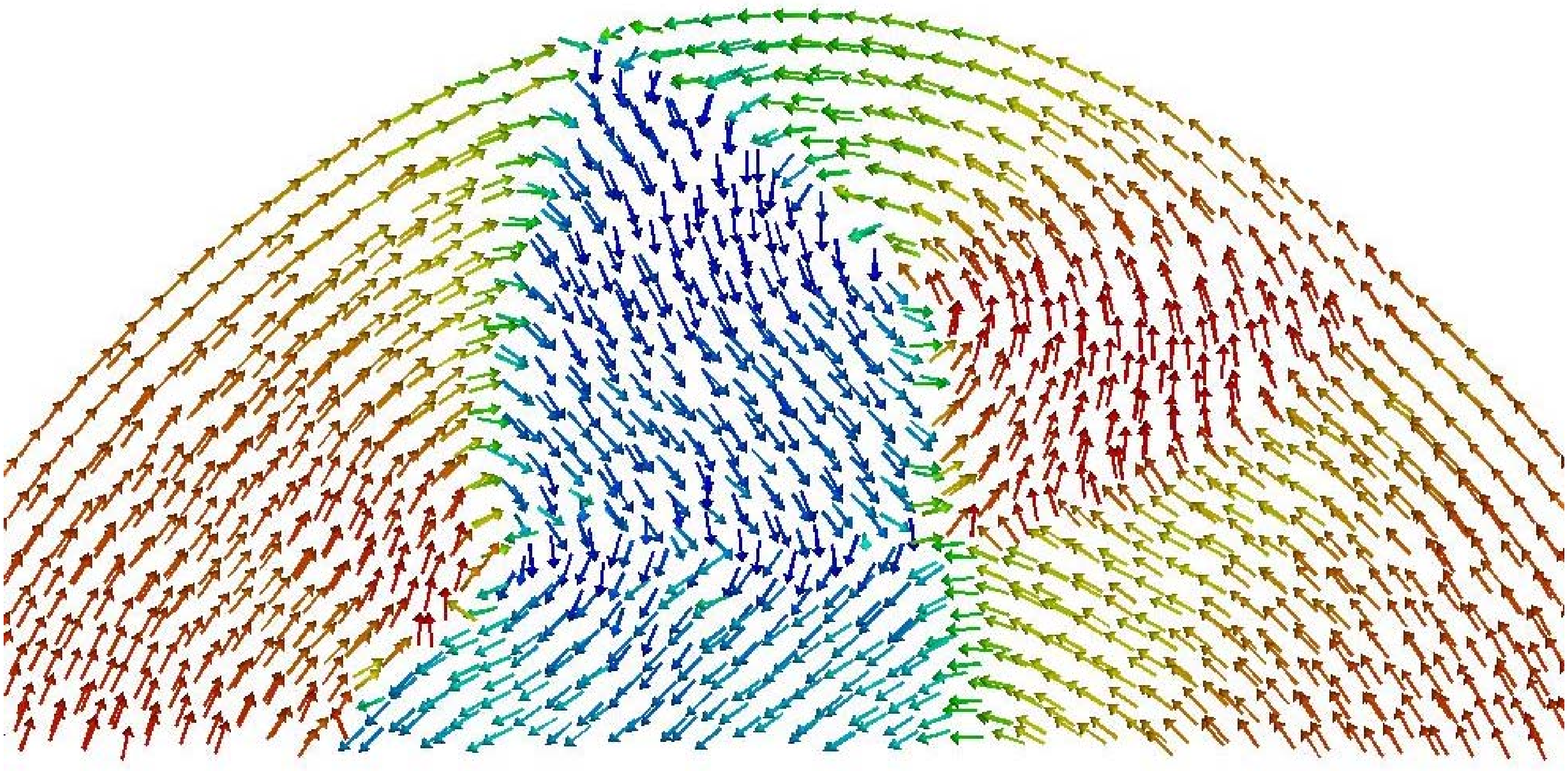}
\caption{More detailed magnetisation configuration in the upper part of the ellipsoid showing the occurrence of domain walls and vortices. $H_{dc}=-31.41Oe$, $H_{ac}= 25.13Oe$, $\nu=0.1 GHz$.}
\label{Vortex}
\end{figure}

Generally speaking, the ac-driven process is complicated and involves different objects. The domains are separated by domain walls (of N\'eel or cross-tie types) and the magnetic moments in the center of these domain walls are constantly precessing (see Fig.~\ref{MW} (c) and (d)). This precession is important for the domain wall mobility. In the junctions between the domains vortices are created (see Fig.~\ref{Vortex}).
Next, during the process we observe constant spinwave generation and their reflection from the ellipsoid boundaries and domain walls. Therefore, the overall process contains an interplay of several important effects, discussed earlier in the literature in more simple situations (see the Introduction). Each of them has been reported previously to be influenced by the mw-field.

\begin{figure}[h!]
\includegraphics[height=6cm]{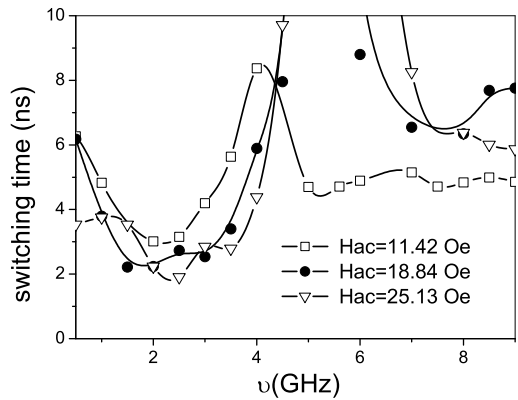}
\caption{Switching time of the magnetisation as a function of the mw frequency for various values of the mw-field amplitude $H_{ac}$ and  $H_{dc}= -31.41 Oe$ }
\label{time}
\end{figure}

\begin{figure}[h!]
\includegraphics[height=6.5cm]{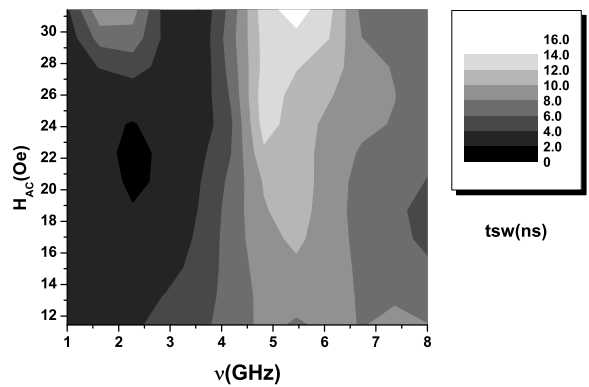}
\caption{Contour plot for the switching time of the ellipsoid for applied field $H_{dc}=- 31.41 Oe$. }
\label{contur}
\end{figure}

\begin{figure}[h!]
\includegraphics[height=6cm]{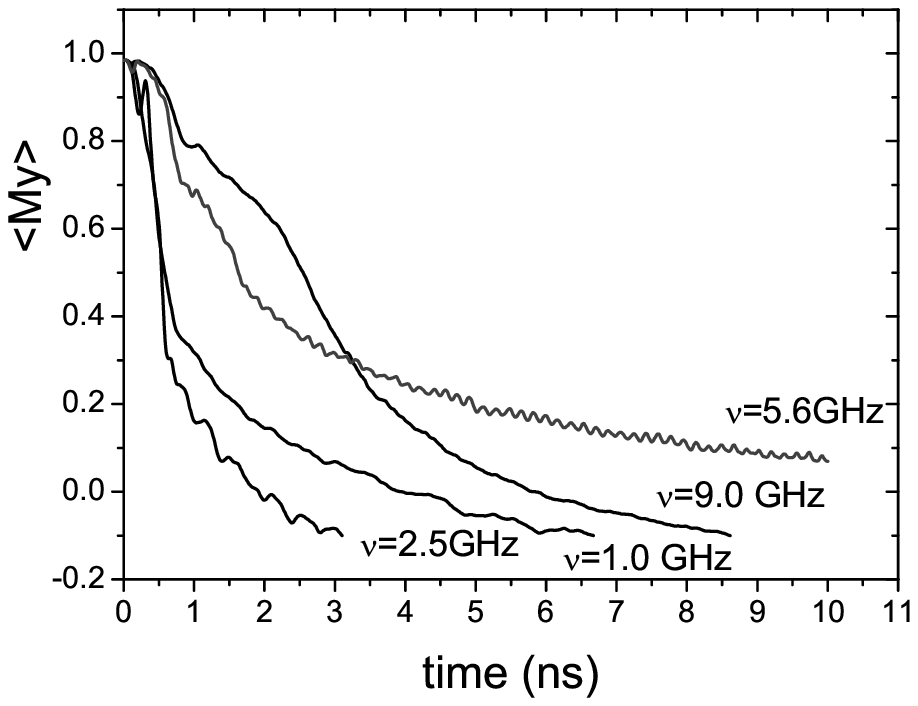}
\caption{Temporal evolution for the average $<M_y>$ magnetisation component (normalized to the saturation value) during the microwave-assisted switching process at $H_{dc}=- 31.41 \; Oe$ and $H_{ac}= 25.13\; Oe$. }
\label{My}
\end{figure}

Fig.~\ref{time} shows the dependence of the magnetisation switching time in the ellipsoid on the mw frequency for several amplitudes of the ac field. We clearly observe  oscillations in the switching time. Moreover there exists a region of the parameters (see Fig.~\ref{contur}) where the switching required more than $16 \;ns$. To understand the phenomenon we first plot in Fig. \ref{My} the average $M_y$ magnetisation component as a function of time. The small-period magnetisation oscillations are related to the ac-field oscillations. During the switching process we can distinguish two main processes: The first one (occurring in the first 2 ns) is related to the efficiency of the magnetisation nucleation in the system while the second one is related to the magnetisation relaxation. Fig.~\ref{My} shows that at several frequencies the magnetisation reversal process started faster in its nucleation part but proceeded slower in the relaxation part.

 The nucleation process is associated to efficient spin wave generation and then to  spinwave instability phenomena which is a precursor of the magnetisation reversal \cite{Chubykalo, Dobin, Kashuba}. The role of the ac-field is to excite a spinwave with the external frequency.  Because of this, the data of Fig.~\ref{ripple} resembles the spinwave dispersion relation and the ripple configurations - the spinwave modes reported in Ref.~\onlinecite{Gubbiotti}. However, the excited mode may be unstable for a given condition and, thus, produce instabilities and magnetisation reversal. For example, we have seen that at these values of the applied field the main FMR mode is completely unstable. When we tried to excite the homogeneous mode, with the magnetisation anti-parallel to the applied field direction, the energy was immediately redistributed into inhomogeneous spinwaves provoking later the magnetisation reversal. The excitation of these spinwaves was acting as an additional damping source leading to the magnetisation reversal similar to the case of Ref.~\onlinecite{Safonov}. In this case, the excitation with the frequencies close to FMR  (around $3 GHz$ in Fig.~\ref{time}) is the most efficient one. The first stage (several ns) is also characterized by initial expansion of the domain size, especially in the center, as seen in Fig.~\ref{size}.

\begin{figure}[h!]
\includegraphics[height=6cm]{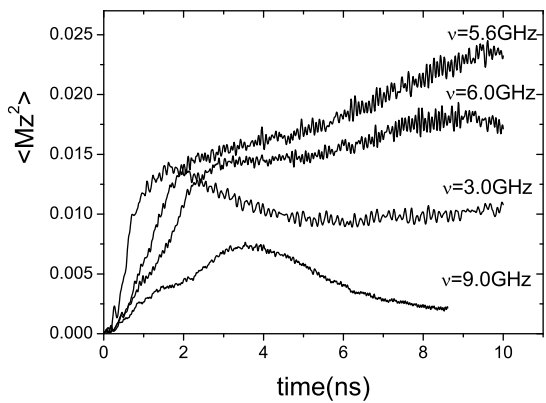}
\caption{Temporal evolution for the average $<M_z^2>$ magnetisation component (normalized to the saturation value) during the microwave-assisted switching process at $H_{dc}=- 31.41 \; Oe$ and $H_{ac}= 25.13 \; Oe$.}
\label{desv}
\end{figure}

To illustrate the role of precession, we present in Fig.~\ref{desv} the temporal evolution of the average value of the $<M_z^2>$ component (this value is proportional to the energy put into precession by the microwave). The first part of these curves characterizes the initial nucleation-expansion process (first 2 ns). It is clear that the energy is put efficiently into the precession with the frequency close to the main FMR mode. This is confirmed by the fast magnetisation decay of the $<M_y>$ component in Fig.~\ref{My} and the fast growth of the $<M_z^2>$ value in Fig.~\ref{desv}. This fact is consistent with the hypothesis introduced in many papers stating that for efficient switching the mw frequency should coincide with the FMR one \cite{Woltersdorf, Sun}. In the case of magnetic elements behaving as one macrospin magnetic moment this would determine the overall switching. However, a fast nucleation is insufficient for  fast magnetisation switching of larger elements. Fig.~\ref{desv} demonstrates that for frequencies close to $6GHz$ (double FMR frequency in the opposite well) the precessional energy remains constant. We note that only the magnetic moments in the centers of domain walls, separating domains with opposite $M_x$ signs (and in the vortex centers) are precessing (see Fig.~\ref{MW}). Thus the expansion of domain walls necessary for magnetisation relaxation is dependent on the relaxation of these magnetic moments. For frequencies and timescales at which the $<M_z^2>$ relaxation is very slow, the energy is transferred efficiently into the precessional motion and not to the relaxation process. Consequently, the further domain expansion is extremely slow. We would like to note that this idea is in the spirit of the hypothesis introduced in Ref.~\onlinecite{Krasyuk} (self-trapping of magnetic oscillations). At the long-time scale the magnetisation reversal at this stage proceeds again via additional spin-wave generation and reflection destabilizing the domain structure. Note also that the nucleation process occurs  when the magnetisation is anti-parallel to the field direction while in the relaxation part it is parallel. Thus, the relevant frequencies are different in both cases.

\begin{figure}[h!]
\includegraphics[height=6cm]{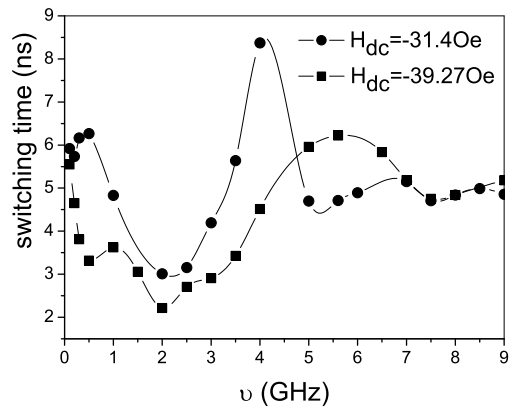}
\caption{Switching time of the magnetisation as a function of the mw frequency for two values of the stationary field $H_{dc}$ and  the mw-field amplitude $H_{ac}= 25.13 \; Oe$.}
\label{Twofields}
\end{figure}

Fig.~\ref{Twofields} shows the magnetisation switching time as a function of the mw frequency for two dc applied field values. As the applied dc field increases, the angle of the magnetisation precession increases in the nucleation part of the process but decreases in the relaxation part. Larger precessional angles lead to nonlinear phenomena. Associated with larger precessional angles there is a nonlinear shift of the frequencies to smaller values \cite{Usatenko}. This is in agreement with the results presented in Fig.~\ref{Twofields}, provided that the relevant frequencies are determined by the relaxation process. However, in Fig.~\ref{time} the frequencies are shifted to larger values with larger ac-field amplitude. This shows that the overall process is much more complicated and cannot be analysed in terms of the FMR frequencies relevant to the switching of one magnetic moment only. In fact, different ac-frequencies excite modes with different spin wave vector and the instabilities of them occur at different threshold amplitudes.

\section{Conclusions} 

Using the micromagnetic model, we have investigated the mechanism of  fast magnetisation switching assisted by a linearly polarised microwave field in micron-sized magnetic elements. Magnetisation dynamics in these magnetic patterns is governed by nucleation-propagation-relaxation processes and is different to those of  nanoscale elements behaving as one macrospin. Our simulations confirm that the microwave-assisted magnetisation reversal requires a smaller field than in the process under a static applied field. The first contribution to this is the deviation of the field from the easy-axis during the mw-assisted switching process. However, even when we compared the situations when the maximum applied field angle is the same in both cases,  mw-assisted switching  appeared to require less field. Moreover, our results show that at small  intensity
of the mw-field its action is not described by the effective field tilt.
 This happens due to the fact that the static and the mw-assisted switching processes involve different reversal modes. In the first case this mode consists of the nucleation of two vortices in  opposite corners, their subsequent merging and creation of two domain walls which later span the whole ellipsoid. In the second case, the reversal mode consists of a ripple structure, the external frequency being responsible for the excitation of a spinwave mode with the corresponding ripple size. For fields below the static coercivity field, the magnetisation configuration corresponds to almost homogeneous background plus  excited spin wave mode and is metastable. The amplitude of the  excited spin wave mode grows with applied field  and becomes unstable, leading to  magnetisation reversal. Thus, the spin-wave instability mechanism \cite{Chubykalo, Dobin,Kashuba} plays a major role for the fast magnetisation switching process in nano-size magnetic elements.

We have shown that the most efficient nucleation is at the FMR frequency (and its multiples). However, the magnetisation reversal process after the nucleation requires also an efficient magnetisation relaxation. For example, the domain growth is stimulated by the ac field. This happens due to the relaxation of precessing magnetic moments in the domain wall center. In the spirit of the mechanism suggested in Ref.~\onlinecite{Krasyuk}, for some frequencies, the relaxation process is not efficient, this happens when the mw field is coupled to the precessional motion. In this case the microwave field efficiently puts the energy into precession and not into the propagation-relaxation processes which are slowed down.
As a consequence of the interplay of several mechanisms with different relevant frequencies, the switching time of magnetic elements is a complicated function of the external frequency. The above results show that the magnetisation dynamics in micron-sized magnetic elements could not be analysed as a simple FMR-related phenomenon and neither follows a direct correspondence with a spinwave spectrum.

Finally, we would like to discuss our results in the context of available experimental data.
 Out results are in qualitative agreement with experimental observations. The Kerr images in Ref.~\onlinecite{Pimentel} show the creation of  ripple structures during the mw-assisted switching process in a permalloy ellipsoid. Currently there is no data available for the switching time of the magnetisation in ellipsoids which could confirm the existence of the oscillations presented in Fig.~\ref{time}. However, in several experimental papers on different magnetic elements we have found qualitative similarities with our results. For example, in Co bars the relaxation time observed by the fast Kerr technique showed a strong nonmonotonic behavior as a function of applied field \cite{Adam}. Also in Co bars measured by the anisotropic magnetoresistance effect \cite{Grollier}, the authors have observed the occurrence of several resonance peaks sweeping in frequency. The microwave power input necessary for the maximum coercivity reduction was reported to be frequency-dependent in the measurements of  mw-assisted switching in NiFe magnetic tunneling junctions \cite{Moriyama}. The critical field necessary for the microwave assistance was shown to be a complicated function of the mw frequency with several minima for the mw- assisted process going from vortex to onion state in magnetic nanorings \cite{Podbielski}.

\section*{Acknowledgements}
The results of the German group were obtained with the research funding from the European Commission under EU-RTN ULTRASWITCH (HPRN-CT-2002-00318) project. The Spanish authors acknowledge  financial support from the European COST-P19 Action and  Spanish National projects MAT2007-66719-C03-01 and
 Consolider CS2008-023. They are also grateful to J.M. Rodriguez Puerta for his help with computer facilities.

\end{document}